\documentclass[aps,prb,twocolumn,superscriptaddress]{revtex4-2}
\usepackage{graphicx}
\usepackage{bm}
\usepackage{xcolor}
\usepackage{amsmath}
\usepackage{amssymb}
\usepackage{hyperref}
\usepackage{tabularray}
\usepackage{wasysym}

\begin{document}
	
	\title{Spin excitations arising from anisotropic Dirac spinons in YCu$_3$(OD)$_6$Br$_2$[Br$_{0.33}$(OD)$_{0.67}$]}
	\author{Lankun Han}
	\affiliation{Beijing National Laboratory for Condensed Matter Physics, Institute of Physics, Chinese Academy of Sciences, Beijing 100190, China}
	\affiliation{School of Physical Sciences, University of Chinese Academy of Sciences, Beijing 100190, China}
	\author{Zhenyuan Zeng}
	\affiliation{Beijing National Laboratory for Condensed Matter Physics, Institute of Physics, Chinese Academy of Sciences, Beijing 100190, China}
	\affiliation{School of Physical Sciences, University of Chinese Academy of Sciences, Beijing 100190, China}
	\author{Min Long}
	\affiliation{Department of Physics and HK Institute of Quantum Science \& Technology, The University of Hong Kong, Pokfulam Road, Hong Kong, China}
	\author{Menghan Song}
	\affiliation{Department of Physics and HK Institute of Quantum Science \& Technology, The University of Hong Kong, Pokfulam Road, Hong Kong, China}
	\author{Chengkang Zhou}
	\affiliation{Department of Physics and HK Institute of Quantum Science \& Technology, The University of Hong Kong, Pokfulam Road, Hong Kong, China}
	\author{Bo Liu}
	\affiliation{Beijing National Laboratory for Condensed Matter Physics, Institute of Physics, Chinese Academy of Sciences, Beijing 100190, China}
	\author{Maiko Kofu}
	\affiliation{J-PARC Center, Japan Atomic Energy Agency, Tokai, Japan}
	\author{Kenji Nakajima}
	\email{kenji.nakajima@j-parc.jp}
	\affiliation{J-PARC Center, Japan Atomic Energy Agency, Tokai, Japan}
	\author{Paul Steffens}
	\affiliation{Institut Laue-Langevin, 71 Avenue des Martyrs, 38000 Grenoble, France}
	\author{Arno Hiess}
	\affiliation{Institut Laue-Langevin, 71 Avenue des Martyrs, 38000 Grenoble, France}
	\affiliation{European Spallation Source ERIC, P.O. Box 176, 22100 Lund, Sweden}
	\author{Zi Yang Meng}
	\affiliation{Department of Physics and HK Institute of Quantum Science \& Technology, The University of Hong Kong, Pokfulam Road, Hong Kong, China}
	\author{Yixi Su}
	\email{y.su@fz-juelich.de}
	\affiliation{J\"{u}lich Centre for Neutron Science (JCNS) at Heinz Maier-Leibnitz Zentrum (MLZ), Forschungszentrum J\"{u}lich, Lichtenbergstrasse 1, 85747 Garching, Germany}
	\author{Shiliang Li}
	\email{slli@iphy.ac.cn}
	\affiliation{Beijing National Laboratory for Condensed Matter Physics, Institute of Physics, Chinese Academy of Sciences, Beijing 100190, China}
	\affiliation{School of Physical Sciences, University of Chinese Academy of Sciences, Beijing 100190, China}
	\begin{abstract}
		A Dirac quantum spin liquid hosts Dirac spinons, which are low-energy fractionalized neutral quasiparticles with spin 1/2 that obey the Dirac equation. Recent inelastic neutron scattering studies have revealed a cone spin continuum in YCu$_3$(OD)$_6$Br$_2$[Br$_{x}$(OD)$_{1-x}$], consistent with the convolution of two Dirac spinons. In this work, we further studied spin excitations using the inelastic neutron scattering technique. The width of low-energy spin excitations shows a linear temperature dependence, which can be explained by spinon-spinon interactions with a Dirac dispersion. Polarized neutron scattering measurements reveal that in-plane magnetic fluctuations are about 1.5 times stronger than the out-of-plane ones, suggesting the presence of Dzyaloshinskii-Moriya interaction and consistent with our theoretical modeling and simulations. Moreover, the high-energy spin excitations around 14 meV agree with the one-pair spinon-antispinon excitations in Raman studies. The real part of the dynamical susceptibility derived from the Kramers-Kronig relationship also agrees with the Knight shift measured by nuclear magnetic resonance, clearly demonstrating the negligible effects of magnetic impurities on static susceptibility. These results provide a rare example in studying quantum-spin-liquid materials where different experimental techniques can be directly compared, and they give further insights for the possible Dirac quantum spin liquid in this system.
	\end{abstract}
	
	\maketitle
	
	\section{Introduction}
	
	A Dirac spinon is a kind of fractionalized excitation in quantum spin liquids (QSLs) that has been widely discussed in theories \cite{SavaryL17,ZhouY17,BroholmC20,maDynamical2018,xuMonte2019}. A Dirac spinon has spin 1/2 and is described by the Dirac equation, i.e., it is a fermion with linear dispersion, analogous to electrons in graphene but without the charge degree of freedom. One of the most promising models to study Dirac spinons is the antiferromagnetic (AFM) Heisenberg model on the kagome lattice \cite{HastingsMB00,HermeleM08,HeYC17,ZhuW18,ZhuW19,SongXY19,HeringM19,DupuisE19,SongXY20,IqbalY21,KieseD23,FerrariF24,SeifertUFP24}. Despite theoretical triumphs, the search for Dirac QSLs in kagome materials has largely failed. One of the main reasons may be the existence of disorders and magnetic impurities in these materials \cite{VriesMA08,FreedmanDE10,YYHuang2021,HanTH12,FuM15,HanTH16,KhuntiaP20,FengZL17,WeiYuan2017,FengZL18b,YingFu2021,WeiY21,NormanMR16}, which may blur the sharp features of low-energy spin excitations in a Dirac QSL or even destroy it.

	Recently, a new kagome system, YCu$_3$(OH)$_6$Br$_2$[Br$_{1-x}$(OH)$_{x}$] (YCu$_3$-Br), has shown many interesting results that make it a promising QSL candidate \cite{ChenXH20,ZengZ22,LiuJ22,HongX22,LuF22,LiS24,ZengZ24,XuA24,ShivaramBS24,SuetsuguS24b}. Unlike many other kagome QSL candidates, this system shows no sign of magnetic impurities or weakly correlated spins. Heat capacity measurements clearly reveal behaviors expected for a Dirac QSL, including the quadratic temperature dependence of the specific heat at zero field and a linear term under fields \cite{ZengZ22}. More intriguingly, the spin excitations show a cone continuum that may come from the convolution of two Dirac spinons \cite{ZengZ24}. These results show no sign of disorders and impurities, making YCu$_3$-Br the best candidate for the Dirac QSL so far. However, there are also several results that do not favor or even go against the existence of Dirac spinons. Thermal conductivity measurements do not observe a linear term that is supposed to exist in the presence of spinon Fermi surfaces coming from the Dirac spinons under fields \cite{HongX22}. Moreover, magnetic susceptibility shows almost no temperature dependence at low temperatures \cite{ZengZ22,LiS24,SuetsuguS24}, whereas a linear temperature dependence is expected for a Dirac QSL. These results cast a shadow over the existence of Dirac spinons, and they have prompted alternate explanations \cite{SuetsuguS24b}. It is worth noting that a one-ninth magnetization plateau and magnetic oscillations have been observed at about 20 T, and their origin is also debated \cite{JeonS24,ZhengG23,SuetsuguS24,ZhengG24}. These debates necessitate systematic quantitative cross-validation between different experimental techniques -- a crucial step that remains largely unexplored in existing QSL studies~\cite{mengPerspective25}.
	
	In this work, we studied the spin excitations of YCu$_3$(OD)$_6$Br$_2$[Br$_{0.33}$(OD)$_{0.67}$] using the inelastic neutron scattering (INS) technique. The temperature dependence and anisotropy of the low-energy spin excitations suggest finite quasiparticle lifetimes and the existence of Dzyaloshinskii-Moriya (DM) interaction. Our theoretical modeling of the material and the Landau-Lifshitz dynamics and tensor-network simulations provide consistent results with the experimental spectra. We argue that these factors explain the negative results from magnetic susceptibility measurements. Moreover, the dynamical susceptibility is in excellent agreement with Raman and NMR results, providing further evidence of the Dirac QSL in this system.
	
	\section{Experiments and Theoretical Computation Methods}
	\label{sec:II}
\begin{figure}[tbp]
		\includegraphics[width=\columnwidth]{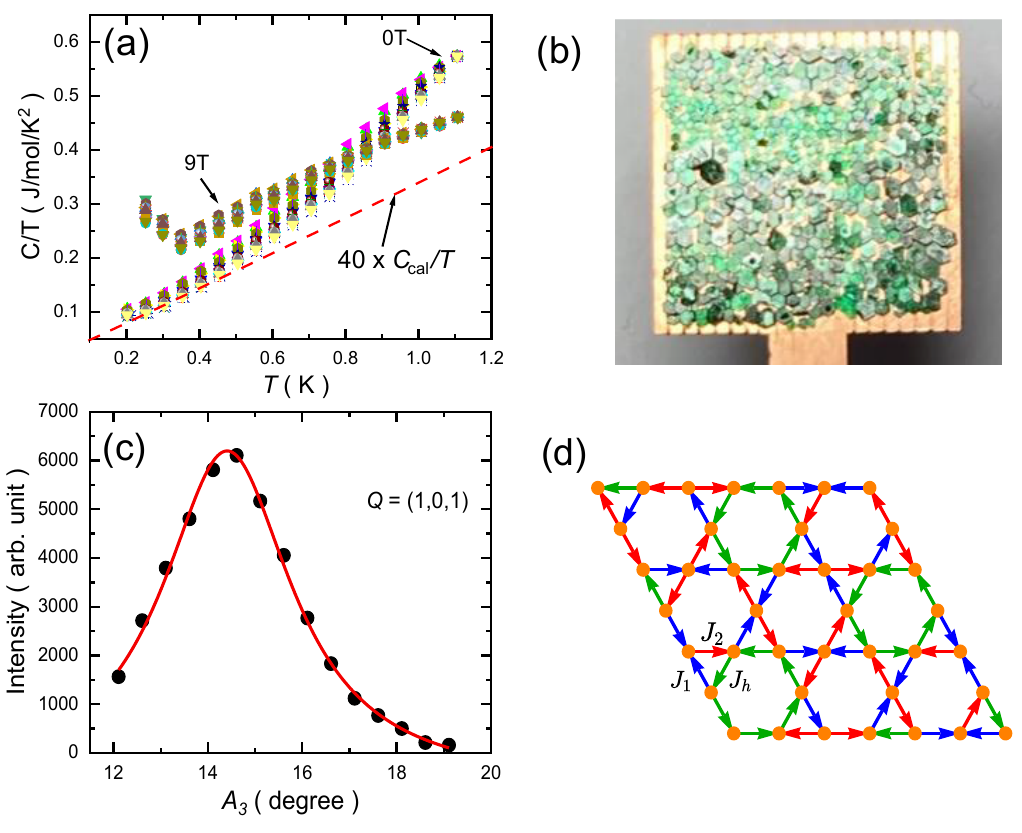}
		\caption{(a) Specific heat below 1 K at 0 and 9 T from 21 different batches of YCu$_3$-Br samples. The upturn below about 0.3 K at 9 T is due to the nuclear Schottky anomaly \cite{ZengZ22}. The red dashed line represents 40 times the specific-heat values calculated by the linear spin-wave theory. (b) A photo of about 800 co-aligned single crystals attached on both sides of a Cu plate. (c) The rocking curves at the (1,0,1) Bragg peak. The solid line is the fitted result of a Gaussian function. (d) Schematic diagram of the 3$J$ Hamiltonian on the kagome lattice including three different couplings: $J$ (depicted by the blue line), $J'$ (the red line), and $J_{\hexagon}$ (the green line). The arrows indicate the definition of the bond orientations $\langle \overrightarrow{i, j}\rangle$ that determine the sign of the Dzyaloshinskii-Moriya interactions in the Hamiltonian.}
		\label{FigExp}
	\end{figure}

	Single crystals of YCu$_3$(OD)$_6$Br$_2$[Br$_{0.33}$(OD)$_{0.66}$] were grown using the hydrothermal method as reported previously \cite{ZengZ22}. Since the sample properties may vary among different batches \cite{XuA24}, we have measured the low-temperature specific heats of the samples from all batches to make sure that the crystals used in the inelastic neutron scattering (INS) measurements exhibit the same properties, as shown in Fig. \ref{FigExp}(a). About 800 single crystals with a mass of approximately 0.5 grams were co-aligned on a Cu plate using CYTOP [Fig. \ref{FigExp}(b)], with a mosaic spread of about 3 degrees, as shown in Fig. \ref{FigExp}(c). The momentum transfer $Q$ in three-dimensional reciprocal space is defined as $Q$ = $H${\bf a$^*$}+$K${\bf b$^*$}+$L${\bf c$^*$}, where $H$, $K$, and $L$ are Miller indices and {\bf a$^*$} = 2$\pi$({\bf b}$\times${\bf c})/V, {\bf b$^*$} = 2$\pi$({\bf c}$\times${\bf a})/V, and {\bf c$^*$} = 2$\pi$({\bf a}$\times${\bf b})/V, with a = b = 6.6779 \AA, c = 5.9874 \AA, and V = 231.23 \AA$^3$ of the kagome lattice. The crystal assembly was aligned in the [$H$, 0, $L$] scattering plane. The unpolarized and polarized INS experiments were carried out on the cold-neutron disk chopper spectrometer AMATERAS at J-PARC \cite{Amateras} and the cold-neutron triple-axis spectrometer ThALES at ILL \cite{Boehm08}, respectively. All the spectral data of AMATERAS were processed by using the software suite Utsusemi \cite{Utsusemi}. In the polarized neutron experiment, an XYZ coil system is used for providing the needed guide fields at the sample, and the incoming neutron polarization directions are denoted as $x$, $y$, and $z$ with $x$ along the $Q$ direction. While both $y$ and $z$ are perpendicular to $Q$, they are within and perpendicular to the scattering plane, respectively. After scattered by the sample, the outgoing neutron polarization can be either parallel or antiparallel to the incoming neutron polarization, which corresponds to the non-spin-flip (NSF) and spin-flip (SF) processes, respectively. 
	
	We also performed theoretical modeling and employed the spin wave and the Landau-Lifshitz dynamics (LLD) simulation~\cite{DahlbomD2022b,DahlbomD2022a,ZhangH2021} as well as the tensor network calculations including the density matrix renormalization group (DMRG)~\cite{white1992dmrg} and the time-dependent variational principle (TDVP)~\cite{Haegeman2011time,Haegeman2016Unify}. Our LLD calculations are performed using Su(n)ny suite, an open-source code developed for simulating equilibrium and non-equilibrium magnetic phenomena from microscopic models~\cite{Sunny2025,DahlbomD2022b,DahlbomD2022a,ZhangH2021}. The modeling of the material is based on the 3$J$ model, referring to the fact that we have three different antiferromagnetic couplings ($J$, $J'$, and $J_{\hexagon}$) on the kagome lattice, as shown in Fig.~\ref{FigExp}(d). The Hamiltonian is written as
	\begin{equation}
		H=\sum_{\langle i,j \rangle}J_{i,j} \mathbf{S}_i \cdot \mathbf{S}_j+ \sum_{\langle i,j \rangle} \mathbf{D}_{i,j}\cdot(\mathbf{S}_i \times \mathbf{S}_j)
		\label{eqH}
	\end{equation}
	where $J_{i,j}$ denotes the nearest coupling given by $J$, $J'$, and $J_{\hexagon}$ based on previous theoretical analyses of similar materials~\cite{HeringM22}, and $\mathbf{D}_{i,j}=(\Delta D, \Delta D, D)$ represents the Dzyaloshinskii-Moriya (DM) interaction \cite{FerrariF23,ZhuW19}. We introduce anisotropy to the DM interaction by adjusting $\Delta$. 
	For tensor network calculations, we work on a finite $L_x \times L_y = 3 \times 6$ cylinder. The ground state is achieved using DMRG. Then we perform TDVP to compute the spectrum. More details of the DMRG and TDVP simulations are given in Sec.~\ref{sec:IV}. 
	
	\begin{figure}[tbp]
		\includegraphics[width=\columnwidth]{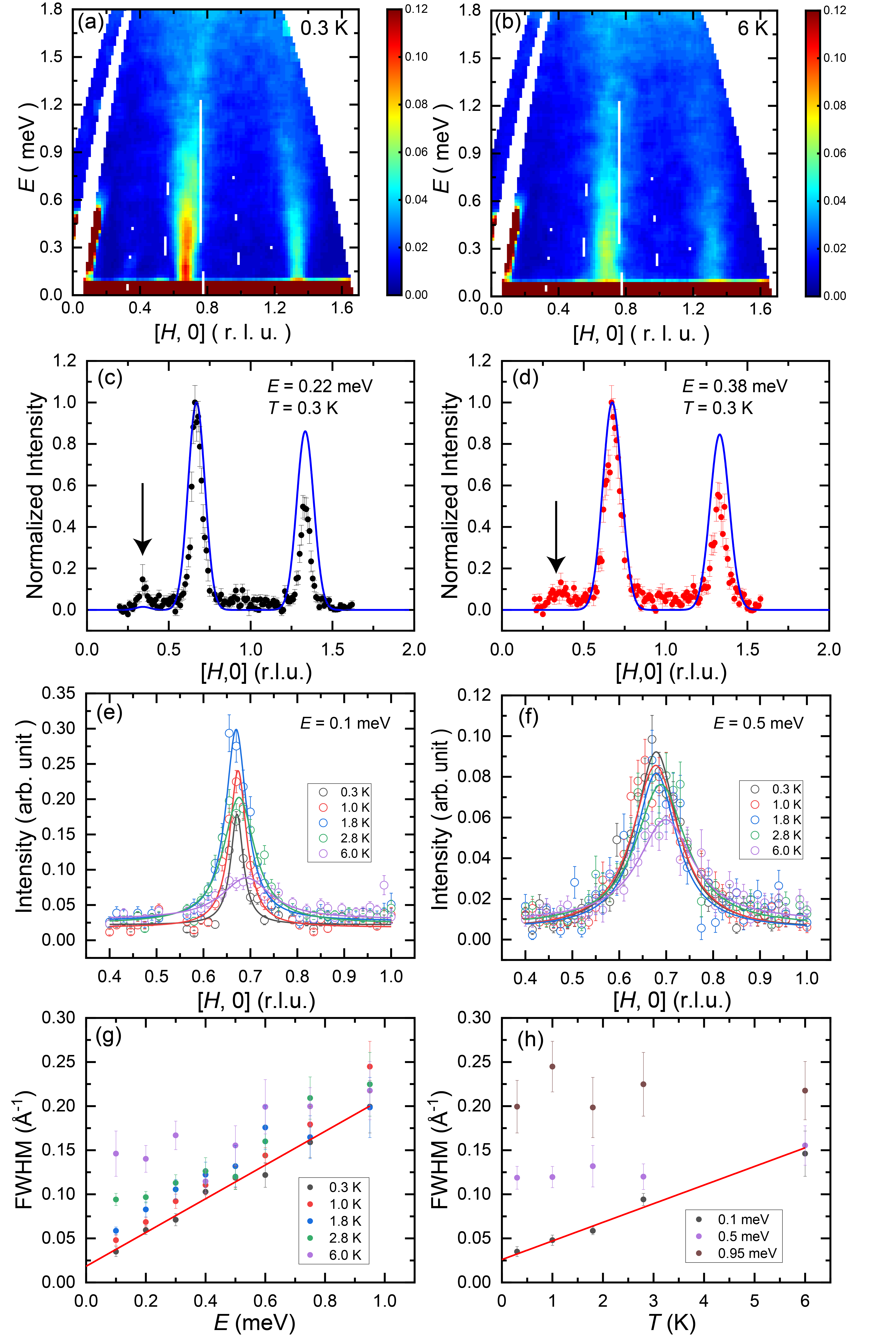}
		\caption{(a) and (b) Intensity contour plots of the INS results as a function of $E$ and $Q$ along the [$H$,0] direction at 0.3 and 6 K, respectively, with $E_i$ = 2.57 meV. (c) and (d) Constant-$E$ cuts along the [$H$,0] direction at 0.22 and 0.38 meV, respectively, at 0.3 K. The solid lines are theoretical calculated results from Ref. \cite{ChatterjeeD23}. The intensities have been normalized to the peak value at (2/3,0). The arrows indicate the (1/3,0) position. (e) and (f) Constant-$E$ cuts along the [$H$,0] direction at various temperatures at 0.1 and 0.5 meV, respectively. The solid lines are fitted results by the Lorentzian function. The integrated ranges of the energy and [-0.5$K$, $K$] are $\pm$ 0.05 meV and $\pm$ 0.1 r.l.u., respectively. (g) and (h) Energy and temperature dependence of the FWHM, respectively. The solid line is the linear-fit result at 0.3 K in (g) and 0.1 meV in (h). In all figures, r.l.u. is the reciprocal lattice unit.}
		\label{LowE}
	\end{figure}
	
	\section{Spin Excitations}
	
	As shown in our previous work \cite{ZengZ24}, the spin excitations of YCu$_3$-Br exhibit cone continua at six symmetrical positions within one Brillouin zone, among which the one at (2/3,0) has the strongest intensity. Figures \ref{LowE}(a) and \ref{LowE}(b) show the colormaps of the low-energy spin excitations as a function of $E$ and $Q$ along the [$H$,0] direction at 0.3 and 6 K, respectively. Similar to previous results, the cone continuum is clearly revealed at both (2/3,0) and (4/3,0) at 0.3 K. At 6 K, the cone continuum becomes column-like due to the significant broadening of low-energy spin excitations. It is interesting to compare our results with previous theoretical calculations for damped magons in Y$_3$Cu$_9$(OH)$_{19}$Cl$_8$ (Y$_3$Cu$_9$-Cl) based on the 3J model with significant randomness \cite{ChatterjeeD23}, as shown in Figs. \ref{LowE}(c) and \ref{LowE}(d). An inexplicable experimental result is the significant intensity at (1/3,0), which is forbidden by spin-wave theory yet matches the structure factor of randomly arranged nearest-neighbor singlets \cite{ZengZ24}. While randomness and disorders can yield weak intensity at (1/3,0) according to  calculations \cite{ChatterjeeD23}, the calculated values are substantially smaller than experimental observations.

	\begin{figure}[tbp]
		\includegraphics[width=\columnwidth]{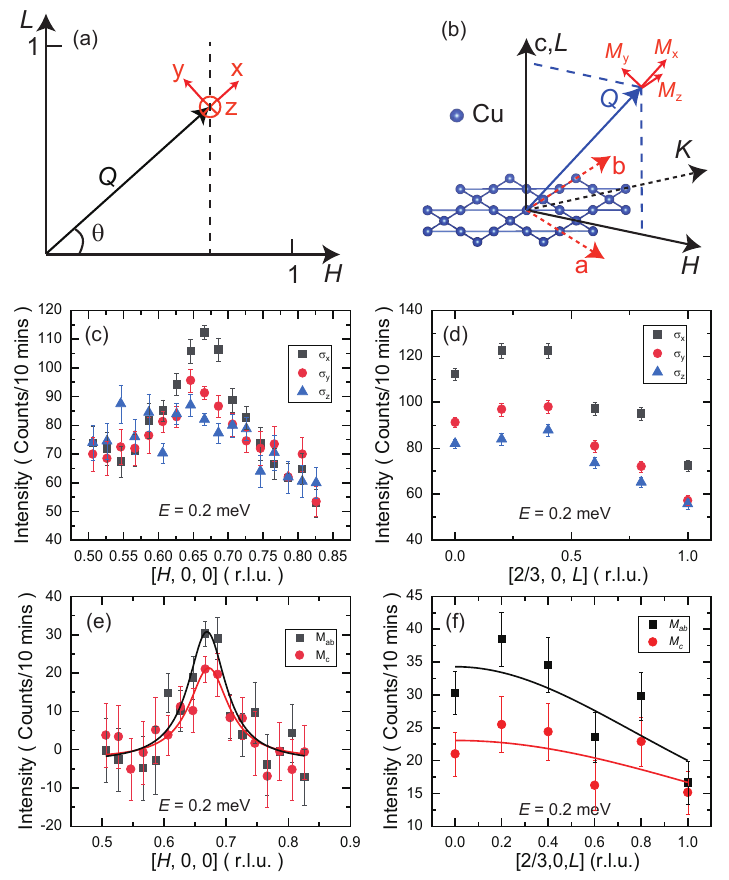}
		\caption{(a) and (b) Illustration of the neutron polarized measurement in the [$H$,0,$L$] plane and real space. $x$, $y$, and $z$ are the neutron polarization directions. The angle between $Q$ and the ($H$,0,0) direction is denoted as $\theta$. (c), (d) Constant-$E$ scans at 0.2 meV along the [$H$,0,0] and [2/3,0,$L$] directions, respectively. (e), (f) In-plane and out-of-plane magnetic responses ($M_{ab}$ and $M_c$) at 0.2 meV along the [$H$,0,0] and [2/3,0,$L$] directions, respectively. The solid lines in (e) are fitted by a Lorentzian function. The solid lines in (f) are guides for the eye. All measurements in this figure were carried out at about 50 mK and with a fixed $k_f=1.3$ \AA$^{-1}$ .}
		\label{polarize}
	\end{figure}
	
	Figures \ref{LowE}(e) and \ref{LowE}(f) display constant-$E$ cuts at 0.1 and 0.5 meV, respectively, all of which can be well fitted by the Lorentzian function. The obtained energy and temperature dependence of the full width at half maximum (FWHM) is shown in Figs. \ref{LowE}(h) and \ref{LowE}(i), respectively. Consistent with previous results, the FWHM changes linearly with $E$ at low temperatures. The slope at 0.3 K is 0.191 $\pm$ 0.011 eV\AA, which is larger than the value reported previously \cite{ZengZ24}. This agrees with the fact that previous INS measurements used samples with varying $T^2$ coefficients in the specific heat \cite{XuA24}, which are directly associated with the widths of the spin excitations for a Dirac QSL \cite{ZengZ24}. Interestingly, the FWHM at 0.1 meV changes linearly with $T$ up to 6 K, with a slope of about 0.0211 $\pm$ 0.003 \AA$^{-1}$/K, while those above 0.5 meV show little temperature dependence. This behavior may result from the spinon-spinon interaction near Dirac points. It is well known that the inverse inelastic quasiparticle lifetime $\tau^{-1}$ near the Dirac nodes in the intrinsic undoped graphene is linear with both temperature and energy due to electron-electron interaction \cite{LiQ13}. We argue that the same mechanism may also work for Dirac spinons, so the energy width $\Delta E$ of the Dirac dispersion is linear with both temperature and energy, too, as $\Delta E \sim \tau^{-1}\hbar/2$. Because INS experiments detect two-spinon excitations, the momentum width $\Delta Q$ at low energies is also linear with temperature as $\Delta E \sim \hbar\nu_F\Delta Q$, where $\nu_F$ is the spinon velocity.

	\begin{figure}[tbp]
		\includegraphics[width=\columnwidth]{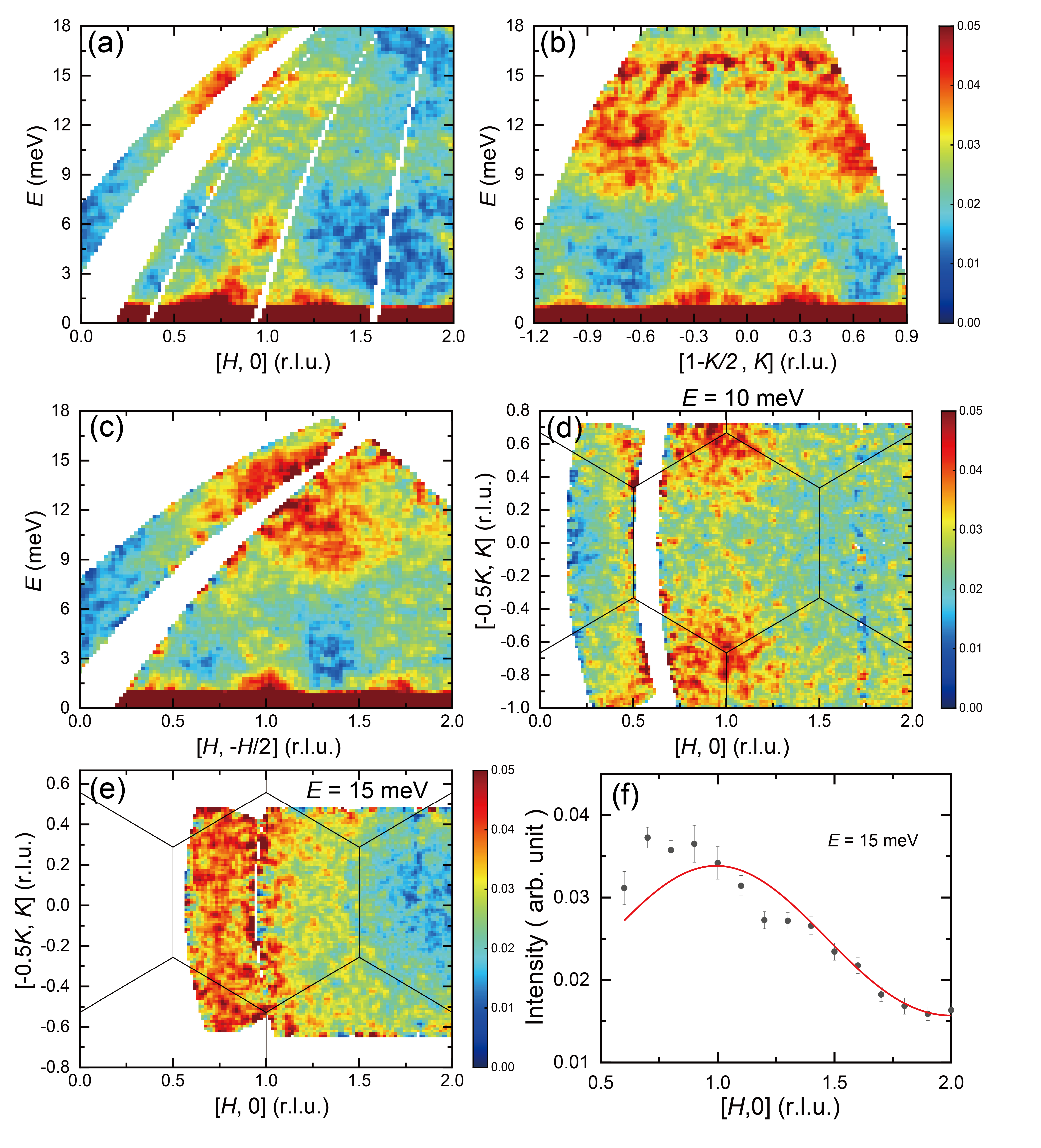}
		\caption{(a)-(c) Intensity contour plots of the INS results as a function of $E$ and $Q$ along the [$H$,0], [1-$K$/2,$K$], and [$H$,-$H$/2] directions, respectively, with $E_i$ = 21.91 meV and $T$ = 0.3 K. The corresponding integrated ranges along the [-$K$/2, $K$/2], [$H$,0], and [0,$K$] directions are from $K$ = -0.12 to $K$ = 0.12, $H$ = 0.9  to $H$ = 1.1, and $K$ = -0.1 to $K$ = 0.1, respectively. (d) and (e) Intensity contour plot of the INS results within the [$H$,$K$] plane at 10 and 15 meV, respectively, at $T$ = 0.3 K. The integrated energy range is 2 meV. The solid lines are the kagome Brillouin zone. (f) Constant-$E$ cut along the [$H$,0] direction at 15 meV. The solid line is the fitted result as described in the main text.}
		\label{HighEColormap}
	\end{figure}

	To further reveal the nature of the low-energy spin excitations, we also carried out polarized INS measurements in the [$H$,0,$L$] plane, as illustrated in Fig. \ref{polarize}(a). Since neutrons only detect magnetic scattering components perpendicular to $Q$, we can obtain magnetic responses along the $y$ and $z$ directions, denoted as $M_y$ and $M_z$, respectively. Accordingly \cite{RegnaultLP06}, we obtain
	\begin{eqnarray}
		M_y = \frac{R+1}{R-1}(\sigma_x^{SF}-\sigma_y^{SF}), \\
		M_z = \frac{R+1}{R-1}(\sigma_x^{SF}-\sigma_z^{SF}),
	\end{eqnarray}
	\noindent where $\sigma_x^{SF}$, $\sigma_y^{SF}$, and $\sigma_z^{SF}$ are the SF scattering cross sections for the corresponding neutron polarization direction, and $R$ is the flipping ratio between NSF and SF scattering. $R$ is large ($>$ 15) so the flipping-ratio correction term can be neglected. As shown in Fig. \ref{polarize}(b), for magnetic intensities of excitations within the $ab$ plane ($M_{ab}$) and along the $c$ axis ($M_c$), we obtain $M_z = M_{ab}$ and $M_y = M_{ab}\sin^2\theta+M_c\cos^2\theta$, where $\theta$ is the angle between $Q$ and the ($H$,0,0) direction. Here, we focus on the spin excitations at 0.2 meV and 50 mK around (2/3,0,0). Figures \ref{polarize}(c) and \ref{polarize}(d) show scans along the [$H$,0,0] and [2/3,0,$L$] directions for all three SF channels, respectively. The corresponding $M_{ab}$ and $M_c$ can thus be obtained as shown in Figs. \ref{polarize}(e) and \ref{polarize}(f). In both cases, the ratio of $M_{ab}/M_c$ is very close to 1.5. We note that this anisotropy can not be due to the anisotropic $g$-factor since $g_{ab}/g_c$ is slightly smaller than 1 as shown by the magnetization measurements \cite{ZengZ22,ZhengG23}. 
	
\begin{figure}[tbp]
		\includegraphics[width=\columnwidth]{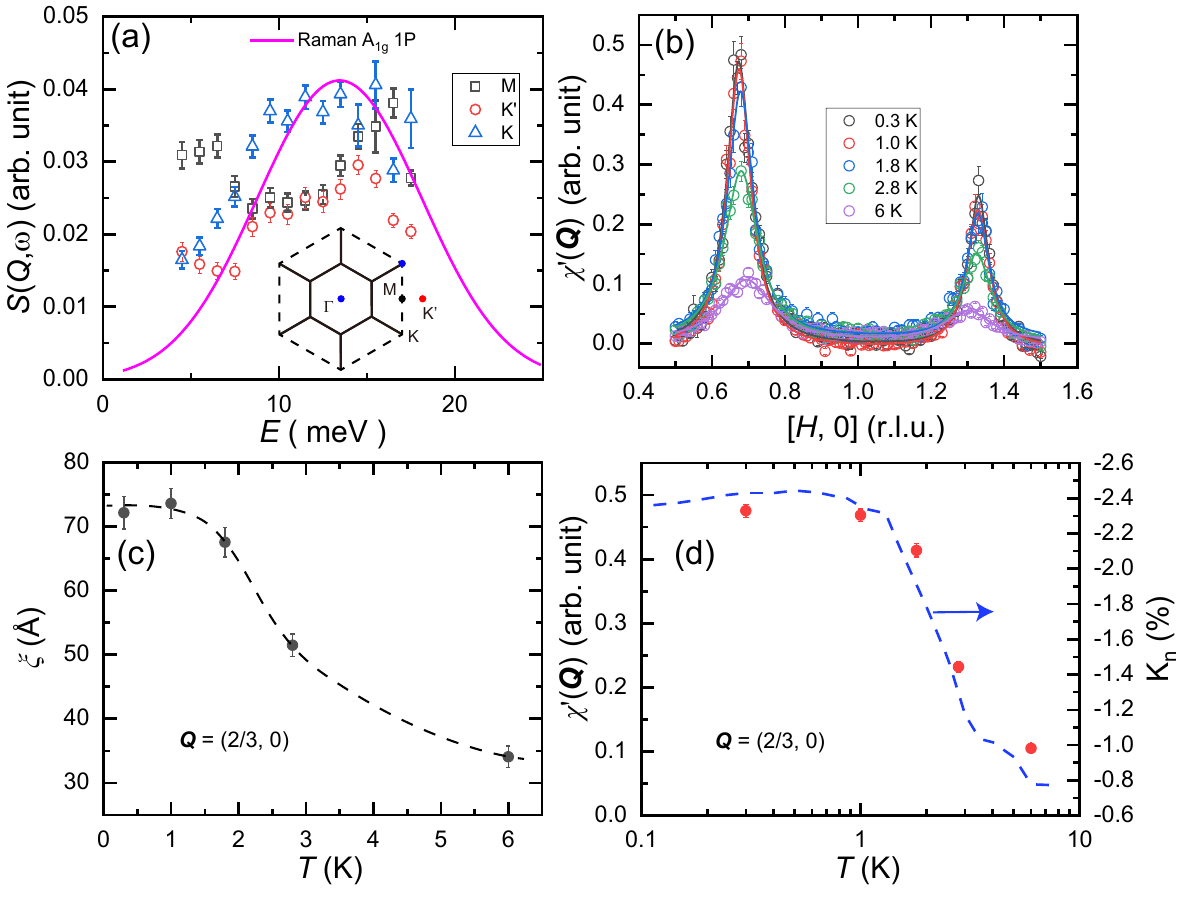}
		\caption{(a) Energy dependence of $S(Q,\omega)$ above 4 meV. The solid line represents Raman magnetic susceptibility for 1P spinon-antispinon excitations in the $A_{1g}$ channel \cite{JeonS24}. The inset shows kagome (solid lines) and extended (dashed lines) Brillouin zones with high-symmetry points. (b) $\chi'(Q)$ along the [$H$,0] direction at several temperatures. The solid lines are fitted results by the Lorentzian function. (c) Temperature dependence of magnetic correlation length $\xi$ at $Q$ = (2/3,0). The dashed line is a guide to the eye. (d) Temperature dependence of $\chi'(Q)$ at $Q$ = (2/3,0). The dashed blue line is the Knight shift from the NMR measurements \cite{LiS24}. }
		\label{HighEAnalyze}
	\end{figure}

	Figures \ref{HighEColormap}(a), \ref{HighEColormap}(b), and \ref{HighEColormap}(c) show high-energy spin excitations at 0.3 K along the [$H$,0], [1-$K$/2,$K$], and [$H$,-$H$/2] directions, respectively. In previous studies, the spin excitations were observed up to about 8 meV due to the incident energy limit \cite{ZengZ24}, while here the excitations have been found up to approximately 18 meV with $E_i$ = 21.91 meV. Similar to previous results \cite{ZengZ24}, the low-energy spin excitations at (2/3,0) and the symmetrical positions gradually merge together at the zone center (1,0) below about 8 meV. At higher energies, new excitations emerge from the zone corners, i.e., (2/3,2/3) and (4/3,-2/3), as further shown in Fig. \ref{HighEColormap}(d). These locations correspond to the $K$ points in the Brillouin zone [see the inset of Fig. \ref{HighEAnalyze}(a)].  When the energy increases to about 15 meV, the spin excitations are observed throughout the Brillouin zone [Fig. \ref{HighEColormap}(e)]. Note that the intensity at 15 meV as a function of $|Q|$ can be approximately fitted by a cosine function, analogous to the structure factor of randomly arranged nearest-neighbour singlets combined with the magnetic form factor of Cu$^{2+}$ \cite{ZengZ24}. The elevated intensities at smaller $Q$ values likely arise from increasing background contributions or stronger intrinsic intensity near the zone boundary.
	
	Previous Raman scattering has revealed high-energy magnetic excitations that are suggested to come from one-pair (1P) and two-pair (2P) spinon-antispinon excitations \cite{JeonS24}. Here, we plot the intensity of our high-energy spin excitations together with the 1P Raman susceptibility in the $A_{1g}$ channel in Fig. \ref{HighEAnalyze}(a). Note that we have combined the neutron data at different temperatures since the high-energy spin excitations are essentially temperature-independent for the temperature range measured here ($T \le$ 6 K). A broad peak is found in neutron scattering data at about 14 meV at the $K$ point, similar to the peak in Raman 1P excitations, suggesting their common origin. Note that the Raman scattering technique detects spinon-antispinon pairs with a total spin quantum number $\Delta S$ = 0, while INS measures two spinon excitations with $\Delta S$ = $\pm$1 \cite{note1}. The probability of two excited spinons with the same and opposite spin quantum numbers could be same so that the results of Raman and neutron scattering are consistent. 
	These excitations also exist at the $M$ and $K'$ points, but are significantly enhanced with decreasing energy compared with Raman susceptibility. Note that Raman spectroscopy only detects signal at $\Gamma$ ($Q$ =0), so its results cannot be directly compared with the low-energy spin excitations measured by the INS technique at finite $Q$'s. 
	
	As shown previously, we can obtain the real part of the dynamical susceptibility $\chi'(Q)$ from the Kramers-Kronig relationship \cite{ZengZ24} as follows,
	\begin{equation}
		\chi'(Q) \propto \int_{-\infty}^{\infty}\frac{\chi''(Q,\omega)}{\omega}d\omega.
		\label{eqKK}
	\end{equation}
	\noindent, where $\chi''(Q,\omega)$ is the imaginary part of the dynamical susceptibility and is calculated as $\chi''(Q,\omega)$ = $S(Q,\omega)(1-e^{-\hbar\omega/k_BT})$. Figure \ref{HighEAnalyze}(b) shows $\chi'(Q)$ along the [$H$,0] direction at various temperatures. The peaks at (2/3,0) and (4/3,0) become much broader and lower at 6 K compared to those at low temperatures. The magnetic correlation length $\xi$ at (2/3,0) can be calculated as $\xi$ = $2\pi/w$, where $w$ is the FWHM of the $\chi'(Q)$ peak at (2/3,0) fitted by the Lorentzian function. Notably, the instrumental resolution is high enough to negligibly affect the determination of $\xi$ \cite{ZengZ24}. Figure \ref{HighEAnalyze}(c) shows the temperature dependence of $\xi$, which increases with decreasing $T$ and saturates below about 1 K. This seems to correspond with the loss of $T^2$ dependence in the specific heat above 1 K \cite{ZengZ22}. Figure \ref{HighEAnalyze}(d) shows the temperature dependence of the peak intensity $\chi'(Q)$ at (2/3,0), which also becomes saturated below about 1 K. Note that the Knight shift $^{81}K_n$ obtained from the nuclear magnetic resonance (NMR) spectroscopy shows similar behavior \cite{LiS24}.

	\section{Theoretical Calculations}
	\label{sec:IV}
	The anisotropic low-energy spin excitations and the high-energy spectra above 8 meV observed in this work provide new information to understand the spin system in YCu$_3$-Br. It is evident that the locations of low-energy spin excitations cannot be explained by the typical Heisenberg model with uniform exchange energies across the small triangles of the kagome lattice. There is increasing evidence that the spin system of YCu$_3$-Br and its siblings should be described by the 3$J$ model~\cite{HeringM22,ChatterjeeD23,LiuJ22,XuA24} as introduced in Sec.~\ref{sec:II}. Although breaking the translational symmetry of the kagome lattice, it successfully reproduces the $Q$ positions of low-energy spin excitations in YCu$_3$-Br \cite{ZengZ24}. Here we present theoretical simulation results based on the 3$J$ model.
	
	For the Hamiltonian in Eq.~\eqref{eqH}, we set $J=J_{\hexagon}=30$ meV and $J'=\alpha J$ with $\alpha<1$ a turning parameter. When $D$ = 0 meV and $\alpha$ is varied, a phase transition happens from the $Q$ = (1/3,1/3) ordered phase to the disordered phase in the classical limit~\cite{HeringM22}. While the linear spin wave (LSW) theory with significant randomness in exchange energies in the ordered phase ($\alpha$ = 0.4) can naively reproduce the low-energy spin excitations \cite{ChatterjeeD23}, it completely fails to describe the high-energy spectra in the experiments, which is expected as YCu$_3$-Br has a magnetically disordered ground state. We thus employ the LLD method, which numerically solves the equation of motion at large-$S$ limit and is able to capture the spin spectrum of the disordered states~\cite{DahlbomD2022b,DahlbomD2022a,ZhangH2021}. This method is known to be an efficient numerical method to sample spin configurations in thermal equilibrium and to simulate the dynamical response of quantum magnets~\cite{HarryL2024,DahlbomD2024,DahlbomD2023}. Successful applications of the method include capturing the temperature evolution of the excitation spectrum of $\mathrm{FeI}_2$\cite{DahlbomD2024}, understanding temperature-dependent SU(3) spin dynamics in the $S = 1$ antiferromagnet $\mathrm{Ba}_2\mathrm{FeSi}_2\mathrm{O}_7$ \cite{Do2023}, and simulating the paramagnetic excitations in $\mathrm{CoI}_2$\cite{Kim2023}.

	\begin{figure}[tbp]
		\includegraphics[width=\columnwidth]{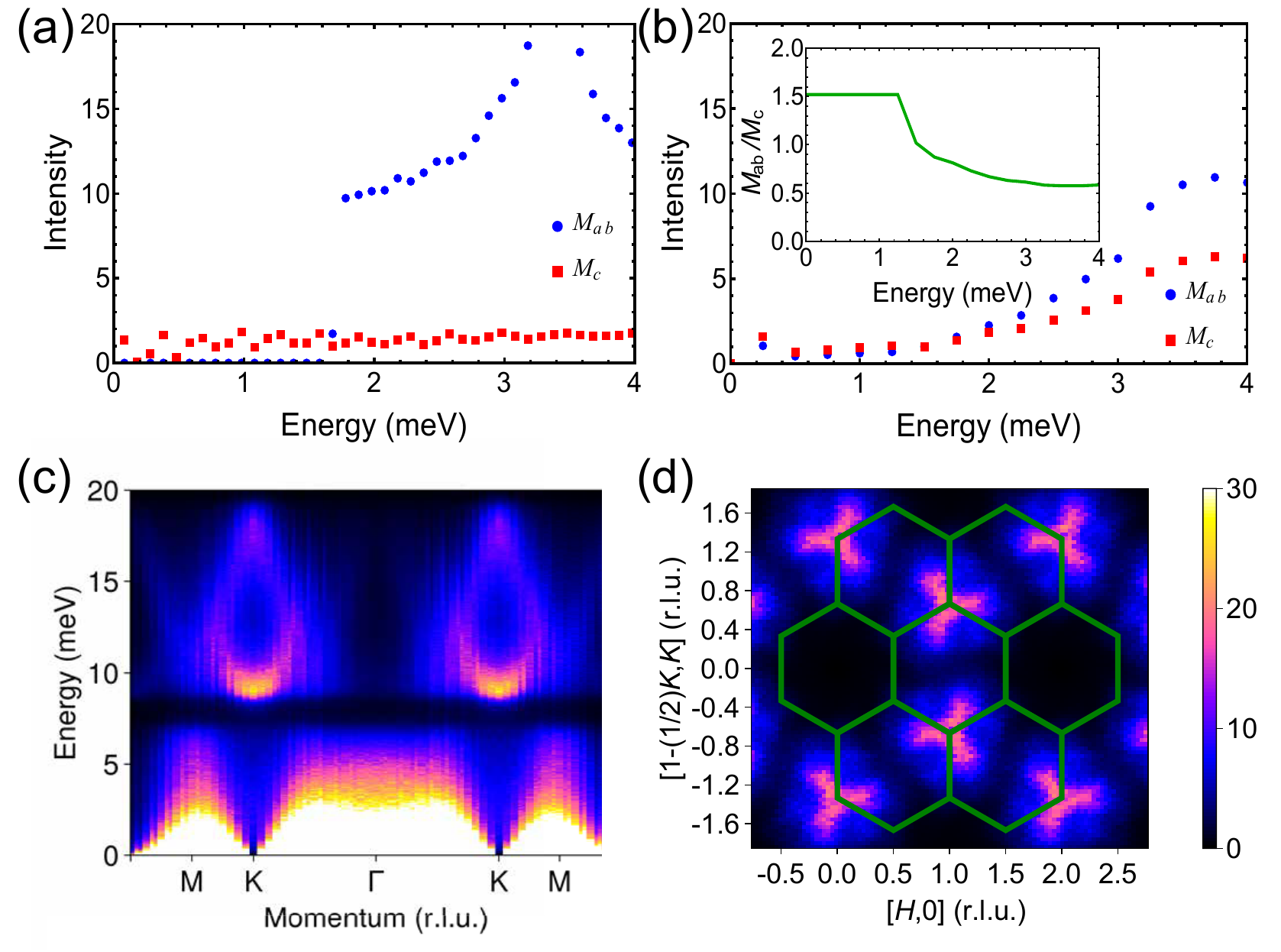}
		\caption{ The LLD results with $J_{\hexagon} = \alpha J = J'/\alpha$ = 30~meV. (a) Energy dependence of calculated $M_{ab}$ and $M_c$ with $Q$ integrated around $K'$ position with $\alpha=0.4$, $D$ = 1.5 meV, and $\Delta$ = 0. (b) Similar to (a) with $\Delta=0.7$. $D$ is assumed to have a randomness from $0.9$~meV to $2.1$~meV. The inset shows the ratio $M_{ab}$/$M_c$.  (c) Contour plots of the LLD calculated spin excitations as a function of energy and momentum along the high-symmetry points, as defined in Fig.~\ref{HighEColormap}(a). Here, $\alpha = 0.6$, $D = 0.0$~meV. (d) In-plane contour plots of the calculated spin excitations at $E = 10$~meV, with solid lines representing the kagome Brillouin zone. }
		\label{LLD}
	\end{figure}
	
	In LLD, a spin configuration is initially established by minimizing the energy of the model on the finite-sized lattice (in our case, a supercell size of $27\times8\times 8$ spins and 100 supercell replicas). This spin configuration is then integrated by the Langevin dynamics that 
	\begin{equation}
		\frac{\mathrm{d}}{\mathrm{d} t}\mathbf{S}_i=-\mathbf{S}_i\times\left[\frac{\partial H}{\partial \mathbf{S}_i}-\lambda\left(\mathbf{S}_i\times\frac{\partial H}{\partial \mathbf{S}_i}\right)+\mathbf{\xi}_i\right],
		\label{eq:eqS2}
	\end{equation}
	where $\mathbf{S}_i$ represents the $i$th spin in the current configuration. The first term on the right hand side ($-\mathbf{S}_i\times\frac{\partial H}{\partial \mathbf{S}_i}$) is derived from the Landau-Lifshitz equation, which provides a classical approximation to the quantum many-body dynamics, neglecting entanglement between sites. And the second term represents the phenomenological Langevin damping, where we have set the damping constant $\lambda=0.1$. $\mathbf{\xi}_i$ denotes Gaussian white noise at each site, which has the following property
	\begin{eqnarray}
		\begin{aligned}
			\langle\xi^{\alpha}_i(t)\rangle&=0\\
			\langle\xi^{\alpha}_i(t)\xi^{\beta}_j(t')\rangle&=2\lambda k_BT\delta_{i,j}\delta_{\alpha,\beta}\delta_{t,t'},
			\label{eq:eqS3}
		\end{aligned}
	\end{eqnarray}
	where $\alpha$ and $\beta$ denote different components of the spins (e.g., $S^x_i$, $S^y_i$, or $S^z_i$).
	
	The spin configuration is then thermalized by running a Langevin dynamic for 10,000 dynamical time steps. This ensures the spin system reaches its low-energy configuration without getting trapped in a local minimum. Subsequently, we progress the spin configurations from thermal equilibrium using with a Langevin trajectory and accumulate the two-spin correlation function $G(r,t)$ during this evaluation. The dynamical time steps in this process should be long enough to decorrelate the spins, and we use 1000 dynamical time steps. Finally, the collected spin correlation function $G(r,t)$ is Fourier transformed to derive the dynamical structure factor $S(Q,\omega)$.
	
	First, we show that the existence of DM interactions, which have been shown to exist in other similar kagome materials, such as herbertsmithite and YCu$_3$(OH)$_6$Cl$_3$ \cite{CepasO08,ZorkoA08,ArhT20}, can naturally explain the anisotropy of low-energy spin excitations. To correctly account for the $Q$ positions of low-energy spectra, the system is chosen as in the $Q$ = (1/3,1/3) ordered state ($\alpha$ = 0.4). Note that the anisotropy caused by the DM interactions is largely independent of whether the ground state is ordered or not. Fig. \ref{LLD}(a) shows the energy dependence of the calculated $M_{ab}$ and $M_c$, where only the out-of-plane DM interaction is considered ($D$ = 1.5 meV and $\Delta$ = 0). A clear energy gap appears in $M_{ab}$ and a similar gap emerges in $M_c$ if only the in-plane DM interaction is taken into account. Introducing randomness will smoothen the gap feature but cannot eliminate it. If both DM interactions are introduced ($D$ = 1.5 meV and $\Delta$ = 0.7), the calculations reproduce the experimentally observed 1.5 ratio of $M_{ab}/M_c$ [Figs.~\ref{polarize}(e) and ~\ref{polarize}(f)], as shown in Fig. \ref{LLD}(b). This clearly demonstrates the role of DM interactions in causing the anisotropic low-energy spin excitations.
	
	\begin{figure}[tbp]
		\includegraphics[width=\columnwidth]{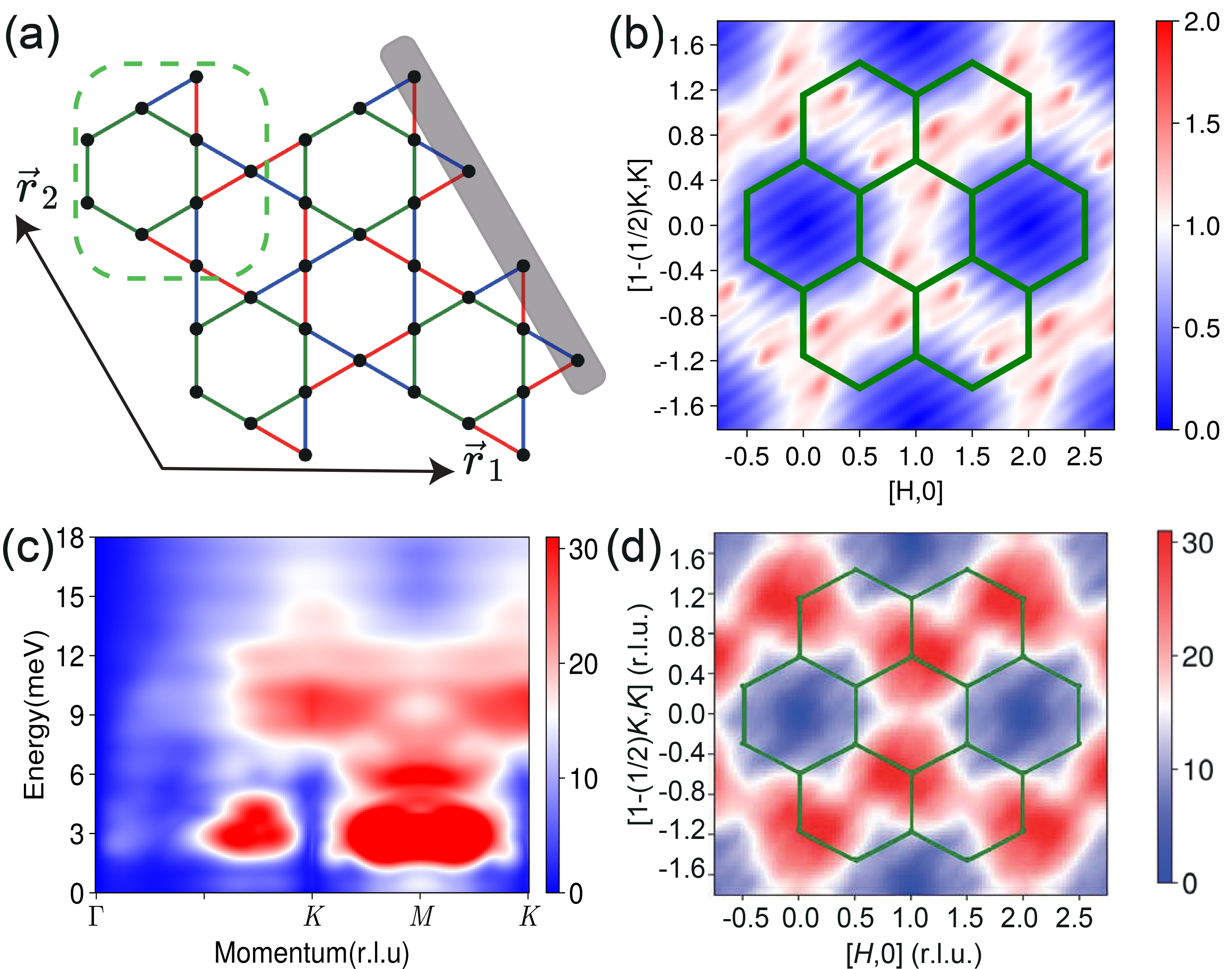}
		\caption{The DMRG results with $\alpha = 0.8$, $D = 0.0$~meV, and $J_{\hexagon} = \alpha J = J'/\alpha = 9$~meV. (a) A sketch of a $2\times 2$ cylinder of 3$J$ kagome model upon which our DMRG calculation is based. The green dashed rectangle denotes a unit cell of the 3$J$ kagome model, with $\vec{r}_1$ and $\vec{r}_2$ its lattice vectors. We impose periodic and open boundary conditions along the direction of $\vec{r}_2$ and  $\vec{r}_1$, respectively. The shaded region is cut to make both cylinder edges of the same armchair shape. (b) The static spin structure  factor $S(Q)$ at $\alpha = 0.8$. The green hexagons depict the kagome Brillouin zone. (c) The spin excitations $S(Q,\omega)$ as a function of energy and momentum along the high-symmetry points, as defined in Fig.\ref{HighEColormap}(a). (d) The spin excitations $S(Q,\omega)$ at $\omega = 10$~meV with solid lines indicating the kagome Brillouin zone.}
		\label{DMRG}
	\end{figure}

	To capture high-energy spectral features in classical spin liquid regime, we perform the LLD simulations with $\alpha$ = 0.6. 
	Fig.~\ref{LLD}(c) shows the spectra along the $M\rightarrow K\rightarrow \Gamma$ path. It reproduces the key features observe experimentally [Figs.~\ref{HighEColormap}(a)-(b)], i.e., significant spectral weight at the $K $-point for $E \geq 9$~meV and at the $\Gamma$-point for 2 meV $\leq E \leq$ 6~meV. Focusing on the $10$~meV mapping of the spectra, Fig. \ref{LLD}(d) demonstrates the spectral weight concentration around the $K$-point, in quantitative agreement with experimental observations in Fig.~\ref{HighEColormap}(c). However, these results requires an unusually large $J_{\hexagon}=30$~meV to reproduce the experimental features, which highlights the quantum nature of the Dirac spin liquid in the material beyond semi-classical treatment in LLD. Therefore, we further perform DMRG calculations to account for the system's quantum nature.
	
	The DMRG calculation is performed on a finite $L_x \times L_y = 3\times6$ cylinder, which is periodic along the $\vec{r}_2$ direction and open along the $\vec{r}_1$ direction, as depicted in Fig.~\ref{DMRG}(a). We further cut off one edge of the open boundary (shaded area in Fig.~\ref{DMRG}(a)) such that both edges of the same shape reduce the edge effect. We first perform the DMRG calculation to obtain the ground state, and then time-evolve it using TDVP with excitation to obtain the spectrum. Since we focus on high-energy spin excitations, the DM interaction $D$ is set to zero. 	
	Spin SU(2) symmetry is implemented in our calculation based on the TensorKit package \cite{Tensorkit_web}.
	We keep the bond dimension up to $D^* = 24000$ states in the DMRG simulation and the maximum truncation error
	below $3 \times 10^{-3}$, where $D^*$ is the number of effective U(1) states kept in our DMRG calculation. In TDVP simulation, we keep up to $D^* = 2000$ states ensuring maximum truncation error below $6\times 10^{-3}$, and time evolve the wave function $N_t = 2000$ steps with $\Delta t = 0.05$, resulting energy resolution $\Delta \omega = \frac{1}
	{N_t \Delta t} = 0.01$ upto
	$\max(\omega) = \frac{1}
	{\Delta t} = 20~\text{meV}$, we set $\hbar = 1$ in the time evolution.
	
	We set the energy unit $J_{\hexagon} = 9~\text{meV}$ to reproduce the experimentally observed feature in our tensor network calculation.  Figure~\ref{DMRG}(b) demonstrates the static spin structure factor $S(Q)=\frac{1}{N}\sum_{\langle i,\alpha,j,\beta\rangle}\mathbf{S}_{i,\alpha}\cdot \mathbf{S}_{j,\beta}\mathrm{e}^{-\mathrm{i}\mathbf{q}\mathbf{(r_{i,\alpha}-r_{j,\beta})}}$ at $\alpha = 0.8$, where $i,j$ label the unit cells and $\alpha,\beta$ label the sublattice. $S(Q)$ is predominantly characterized by peaks at positions resembling those at low energies (considering that one should make three-fold rotation to compare with experimental results).
	
	The spin excitations can be calculated by Fourier transforming the dynamic structure factor
	\begin{equation}
		\begin{aligned}
			\begin{array}{cc}
				S(Q,\omega) =& \frac{1}{\sqrt{N_t}}  \sum_{j,l,\alpha}  e^{i(\omega+i\eta) t_l } e^{-iQ(\mathbf{r_{j,\alpha}-r_{0,\alpha}})}\\ & \big( \langle \mathbf{S}_{j,\alpha}(t_l) \cdot \mathbf{S}_{0,\alpha}\rangle 
				\big)\\
			\end{array}
		\end{aligned}
	\end{equation}
	Here, to compute the time-ordered Green's function $\big( \langle \psi_0 | \mathbf{S}_{j,\alpha}(t_l) \cdot \mathbf{S}_{0,\alpha}|\psi_0\rangle$, we first obtain the Ground state $|\psi_0\rangle$ focusing on the $S = 0$ sector by DMRG, and then we compute $e^{-iHt_l}\mathbf{S}_{0,\beta}|\psi_0\rangle$ by TDVP. We note that $\alpha$ labels the sublattices and $\vec{r}_{0,\alpha}$ is the position $\alpha$ sublattice of a unit cell in the bulk of the cylinder, while $\vec{r}_{j,\alpha}$ runs over all the sublattice of $3\times6=18$ unit cells. $\eta$ is set to $0.05$ to make the integration on time converge.

	Figures \ref{DMRG}(c) and \ref{DMRG}(d) display the DMRG results with the same energy and $Q$ ranges as those in Figs. \ref{LLD}(c) and \ref{LLD}(d) for the LLD calculations. Similar spectra have been observed in both methods,however, DMRG calculations more closely resemble the experimental results. It is especially important to note that LLD requires a large $J_{\hexagon}$ = 30 meV, which is clearly an over estimated value for this system \cite{ZengZ22,JeonS24}. On the other hand, the DMRG calculation uses much smaller exchanges, clearly suggesting strong quantum fluctuations and a highly entangled ground state beyond the classical spin liquid state.

	\section{Discussions}
	
	YCu$_3$-Br has attracted considerable interests due to its potential for hosting a Dirac QSL. The most prominent feature is its low-energy conical spin continuum that may arise from the convolution of two Dirac spinons \cite{ZengZ24}. Here we further show that the FWHM at 0.1 meV changes linearly with temperature [Fig. \ref{LowE}(f)], possibly due to the spinon-spinon interaction, similar to the effect of electron-electron interaction in the intrinsic graphene. Interestingly, the low-energy spin excitations are anisotropic with $M_{ab}/M_{c} \approx$ 1.5, attributable to DM interactions according to our theoretical calculations. Previous studies show that the magnetic susceptibility and Knight shift of YCu$_3$-Br neither decrease linearly with decreasing temperature as $T$ approaches zero nor increase linearly with increasing field at very low temperatures \cite{ZengZ22,LiS24,SuetsuguS24b}, contradicting the expectation for a Dirac QSL where these quantities should be proportional to the density of states. However, this proportionality holds only when spin rotation symmetry is preserved, which is violated in YCu$_3$-Br due to DM interactions. Consequently, magnetic susceptibility or Knight shift cannot serve as effective probes of density of states, since it is well established that in the absence of spin rotation symmetry, magnetic susceptibility approaches a constant at zero temperature, regardless of the nature of low-energy excitations \cite{SavaryL17}. Notably, DM interaction may be crucial for generating a gauge field in this system \cite{KangB24}.
	
	Our results demonstrate direct comparison of INS data against Raman and NMR results in QSL candidates, as Figs. \ref{HighEAnalyze}(a) and \ref{HighEAnalyze}(d) illustrate. Such comparisons are rare in other systems due to practical constraints. The appropriate exchange energies and sharp low-energy excitations in YCu$_3$-Br allow us to obtain almost the entire spectra range and thereby derive the real part of the dynamical susceptibility according to the Kramers-Kronig relationship [Eq. (\ref{eqKK})]. The observed consistency between the Knight shift and $\chi'[(Q=(2/3,0)]$, confirms dominant excitations at (2/3,0) for very low energies, while ruling out magnetic impurities or strong disorders as origins of low-temperature Knight shift and magnetic susceptibility behavior. The consistency between Raman and INS spectra further affirms that spin excitations above 8 meV stem from spinon pairs. Unlike Raman scattering, our INS data reveal $Q$-dependence in high-energy spin excitations emerging from $K$ points of the Brillouin zone, consistent with theoretical calculations for both conventional kagome AFM model \cite{PunkM14,ZhuW19,FerrariF23,HeringM22} and the 3$J$ kagome AFM model in this work. These results establish a coherent picture across experimental techniques and warrant further theoretical studies.
	
	Our theoretical studies show that the 3J model of Eq. \ref{eqH} captures the main features of the dynamical properties of YCu$_3$-Br. Both the LLD [Fig.~\ref{LLD}(d)] and DMRG [Fig.~\ref{DMRG}(d)] calculations demonstrate a spectral weight concentration around the $K$-point, in parallel with the experimental results [Fig.~\ref{HighEColormap}(b)]. Notably, neither approach adequately describes the low-energy conical spin continuum -- a fundamental limitation stemming from LLD's semi-classical nature and DMRG's finite-size constraints. The unusually large $J_{\hexagon}=30$~meV required for LLD further highlights the system's quantum nature. 
	While further quantum many-body study lies beyond the scope of this work, our findings validate the proposed model's capability to capture some key experimental features. This success suggests two promising directions for future theoretical research, i.e., a full dynamical treatment of the Dirac QSL within our framework, and systematic investigations of the model's excitations spectra through advanced numerical approaches~\cite{mengPerspective25}.
	
	While the above results are promising, it is unsurprising that alternate explanations can be proposed, as is common for QSL candidates. The most competitive alternative involves damped magnons, in which long-range order is destroyed by randomness or disorder, but a remnant of spin waves survives, exhibiting continuum-like features. This scenario has been used to explain the spin excitations of Y$_3$Cu$_9$-Cl \cite{ChatterjeeD23}, which shares several features with our observations, such as the $Q$ positions of the low-energy spin excitations and the spin-continuum-like features. 
	Although previous arguments highlighted discrepancies between damped magnons and a true conical spin continuum based on sharpness at very low energies and line shape at higher energies \cite{ZengZ24}, these remain inconclusive. Here we provide more compelling evidence. Calculating the low-temperature specific heat based on the dispersion from the LLD model reveals that the experimental value is approximately 40 times larger than the theoretical prediction, as shown in Fig. \ref{FigExp}(a). Given that the low-temperature specific heat of two-dimensional magnons inversely scales to the square of exchange energies, reconciling this would require reducing the exchange energies by a factor of 6, which will be too small to account for the spin spectra. More crucially, for magnons -- even damped ones -- intensity at $Q$ = (1/3,0) is expected to be negligible due to the structure factor. Yet, in YCu$_3$-Br, significant spectral intensity appears at (1/3,0) [Figs. \ref{LowE}(c) and \ref{LowE}(d)]. The integrated intensity of the peaks matches the structure factor for randomly arranged nearest-neighbor singlets despite sharp low-energy spin excitations indicating long-range correlations \cite{ZengZ24}. We argue that this places a strong constraint on attributing the low-energy spin continuum to damped magnons. Intriguingly, reliable data for spin excitations at (1/3,0) in Y$_3$Cu$_9$-Cl are lacking \cite{ChatterjeeD23}, leaving open the question of whether Dirac spinons could coexist with weak antiferromagnetic order, particularly in the presence of spacial inhomogeneity \cite{XuA24}.

	\section{Conclusions}
	In summary, our systematic study of spin excitations in YCu$_3$-Br provides crucial insights into the possible Dirac QSL in this system. The observed linear temperature dependence of spin excitation broadening reveals quasiparticle interactions characteristic of Dirac spinons, analogous to electron-electron correlations in graphene, while anisotropic magnetic fluctuations ($M_{ab}/M_c \approx 1.5$) are quantitatively explained by Dzyaloshinskii-Moriya interactions that resolve previous discrepancies in susceptibility measurements. Crucially, our work establishes unprecedented experimental cross-validation through comprehensive spectral mapping, particularly the reconciliation of INS and NMR data via Kramers-Kronig analysis, enabled by our complete characterization of the excitation spectrum across all relevant energy scales. Theoretical modeling using the $3J$ Hamiltonian successfully captures key spectral features through both LLD and tensor network calculations, though the necessity for quantum treatments (DMRG/TDVP) over classical approaches underscores the system's strongly quantum-fluctuating nature. Future investigations should focus on developing unified quantum many-body frameworks capable of simultaneously describing both low-energy conical continua and high-energy excitation patterns.

	\acknowledgements
	
	We thank Y. Zhou, K. Jiang, Cristian Batista and Patrick Lee for helpful discussions. This work is supported by the National Key Research and Development Program of China (Grants No. 2022YFA1403400, No. 2021YFA1400400 and No. 2023YFA1406100), the Chinese Academy of Sciences (Grants No. XDB33000000 and No. GJTD-2020-01) and Research Grants Council of Hong
	Kong (Projects Nos. 17309822, HKU C7037-22GF, No. 17302223, and No. 17301924), and the ANR/RGC Joint Research Scheme
	sponsored by RGC of Hong Kong and French National Research Agency (Project No. A HKU703/22). Measurements on AMATERAS were performed based on the approved proposal (Grant No. 2023B0063). Collected data from ThALES (proposal 4-05-901) are available at DOI:10.5291/ILL-DATA.4-05-901.  We acknowledge the HPC2021 system under the Information Technology Services and the Blackbody HPC system at the Department of Physics, University of Hong Kong, as well as the Beijing PARATERA Tech CO.,Ltd. (URL:https://cloud.paratera.com) for providing HPC resources for our LLD and DMRG simulations. 
	
	L.H., Z. Z., and C.Z. contributed equally to this work.


\begin{thebibliography}{73}
\bibitem [{not()}]{note1}%
	\BibitemOpen
	\href@noop {} {}\bibinfo {note} {Strictly speaking, it is not clear whether
		neutron scattering can detect two spinon excitations with a total $S$ = 0 or
		not, but the argument in the text holds in both cases.}\BibitemShut {Stop}%
\end{thebibliography}
\end{document}